\documentclass[preprint2]{aastex}

\slugcomment{Not to appear in Nonlearned J., 45.}

\shorttitle{Earth-like planets albedo Vs continental distribution}
\shortauthors{Sanrom\'a & Pall\'e}

\begin{document}




\title{RECONSTRUCTING THE PHOTOMETRIC LIGHT CURVES OF EARTH AS A PLANET ALONG ITS HISTORY}







\author{E. Sanrom\'a\altaffilmark{1,2} and E. Pall\'e\altaffilmark{1,2}}

\affil{Instituto de Astrof\'isica de Canarias (IAC), V\'ia L\'actea s/n 38200, La Laguna, Spain}

\email{mesr@iac.es}





\altaffiltext{2}{Departamento de Astrof\'isica, Universidad de La Laguna, Spain}






\begin{abstract}

By utilizing satellite-based estimations of the distribution of clouds, we have studied Earth's large-scale cloudiness behavior
according to latitude and surface types (ice, water, vegetation and desert). These empirical relationships are used here to
reconstruct the possible cloud distribution of historical epochs of Earth's history such as the Late Cretaceous (90 Ma ago), 
the Late Triassic (230 Ma ago), the Mississippian (340 Ma ago), and the Late Cambrian (500 Ma ago), when the landmass
distributions were different from today's. With this information, we have been able to simulate the globally-integrated photometric variability of the planet at these epochs. 
We find that our simple model reproduces well the
observed cloud distribution and albedo variability of the modern Earth. Moreover,
the model suggests that the photometric variability of the Earth was probably much larger in past epochs. 
This enhanced photometric variability could improve the chances for the difficult determination of the rotational period and 
the identification of continental landmasses for a distant planets.

\end{abstract}





\keywords{Astrobiology --- atmospheric effects --- Earth --- planets and satellites: atmospheres --- planets and satellites: general --- planets and satellites: surfaces }

\section{INTRODUCTION}

Since the discovery of the first extrasolar planet by \cite{May95}, more than 500 planets orbiting
stars other than the Sun have been detected using a variety of techniques \citep{Bea06, Beu07, Cha07, Udr07}. 
Most of them are gas giants,
as larger planets are easier to detect than smaller, rocky planets. However, tens of planets in
the super-Earth mass range have also been discovered \citep{Bea06, Ben08, Cha09, May09, Que09}. Although we are not yet capable of detecting and exploring planets
like our own (the smallest planet yet known has $R_{P}=0.127R_{J}$; \cite{Bat11}), ambitious ground and space based projects
are already being planned for the next years, suggesting that in the near future it is very likely that Earth-size planets
can be discovered in large numbers \citep{Lin03, Sch09}.

The atmospheric characterization of giant (Jupiter-like) planets has already started with promising 
results (\cite{Tin10}, and references therein), and several atmospheric models are being developed to help us understand these 
observations. In this sense, it is clear that the exploration of our own solar system and its planets will provide a useful 
opportunity for method validation, enabling more accurate determinations, and characterization of extrasolar planets. 
In particular, observations of the solar system rocky planets, including
Earth, will be key for the search for life elsewhere.

In the last years many studies, both observational and theoretical, have been performed to investigate how the Earth 
would look like to an extrasolar observer. One of the observational approaches has been to observe the light reflected by Earth, 
referred to as Earthshine, via the dark side of the moon at visible \citep{Pal03, Qiu03} and near-IR wavelengths \citep{Tur06}. 
Another approach has been the analysis of Earth's light curves from remote sensing platforms \citep{Cow11, Rob11}. Several
authors have also attempted to model the diurnal photometric variability on an Earth-like planet \citep{For01, Tin06a, Tin06b, Fuj10},
while other authors such as \cite{Woo02}, \cite{Arn02, Arn09}, \cite{Sea05} and \cite{Mon05, Mon06} have attempted to
measure the characteristics of the reflected spectrum and the enhancement of Earth's reflectance at 700 nm due to vegetation.
In addition, \cite{Pal08} determined that the light scattered by the Earth as a function of time contains sufficient information, even 
with the presence of clouds, to accurately measure Earth's rotation period. Even crude reconstruction of the continental 
distribution could be attempted given sufficient signal-to-noise observations \citep{Cow09,Oak09}.

However, it is unlikely that, even if we were to find an Earth-twin, that planet will be at a similar stage of 
evolution as the Earth is today. On the contrary, extrasolar planets are expected to exhibit a wide range of ages and 
evolutionary stages. Because of that, it is of interest not only to use our own planet, as it is today, as an exemplar case, 
but also at different epochs.

In this paper, we have used information about the surface properties and continental distribution of the Earth with the aim of 
studying the behavior of the large-scale cloud patterns. We have obtained empirical relationships between the amount of cloudiness 
and their location on Earth's surface depending on latitude and underlying surface types. This relationships have been used to 
reconstruct the possible cloud distribution for different epochs of the Earth such as 90, 230, 340, and 500 Ma ago. With this 
information, we here attempt to reconstruct and understand the photometric variability of these epochs according to their 
different geographic and clouds distribution.

\section{CLOUD AND CONTINENTAL DISTRIBUTION DATA}



%


%


The International Satellite Cloud Climatology Project (ISCCP) was established in the early 1980s. One of its main goals has been to obtain more information about how clouds alter the radiation balance of the Earth. To this end ISCCP has been collecting and analyzing satellite radiance measurements to infer the global distribution of clouds, their properties, and their diurnal, seasonal, and interannual variations.

In our analysis we have used the monthly mean (D2) fractional cloud cover, derived from a combination of both VIS and IR 
measurements, for the 23 year period 1984-2006. The data are given for 2$^\circ$.5x2$^\circ$.5 grid cells in latitude and longitude. The 
ISCCP dataset\footnote{http://isccp.giss.nasa.gov} provides information on several surface, atmospheric and cloud parameters at 
each grid point. In particular total cloud amount, and several cloud types are given at each grid point. For further details on 
the ISCCP data see \cite{Ros96}.

Detailed information on the varying distribution of continents, deserts, vegetation and ice in different epochs of Earth's 
history, is available from Ron Blakey's web page\footnote{http://jan.ucc.nau.edu} on paleoclimate reconstructions. These geological 
maps have been constructed by using interpretations from a wealth of geologic literature and publications, available on 
the project's web page (see, for example, \cite{Fri11}, and references therein).

\begin{figure*}[b]
\begin{center}
 \includegraphics[width=3.3in]{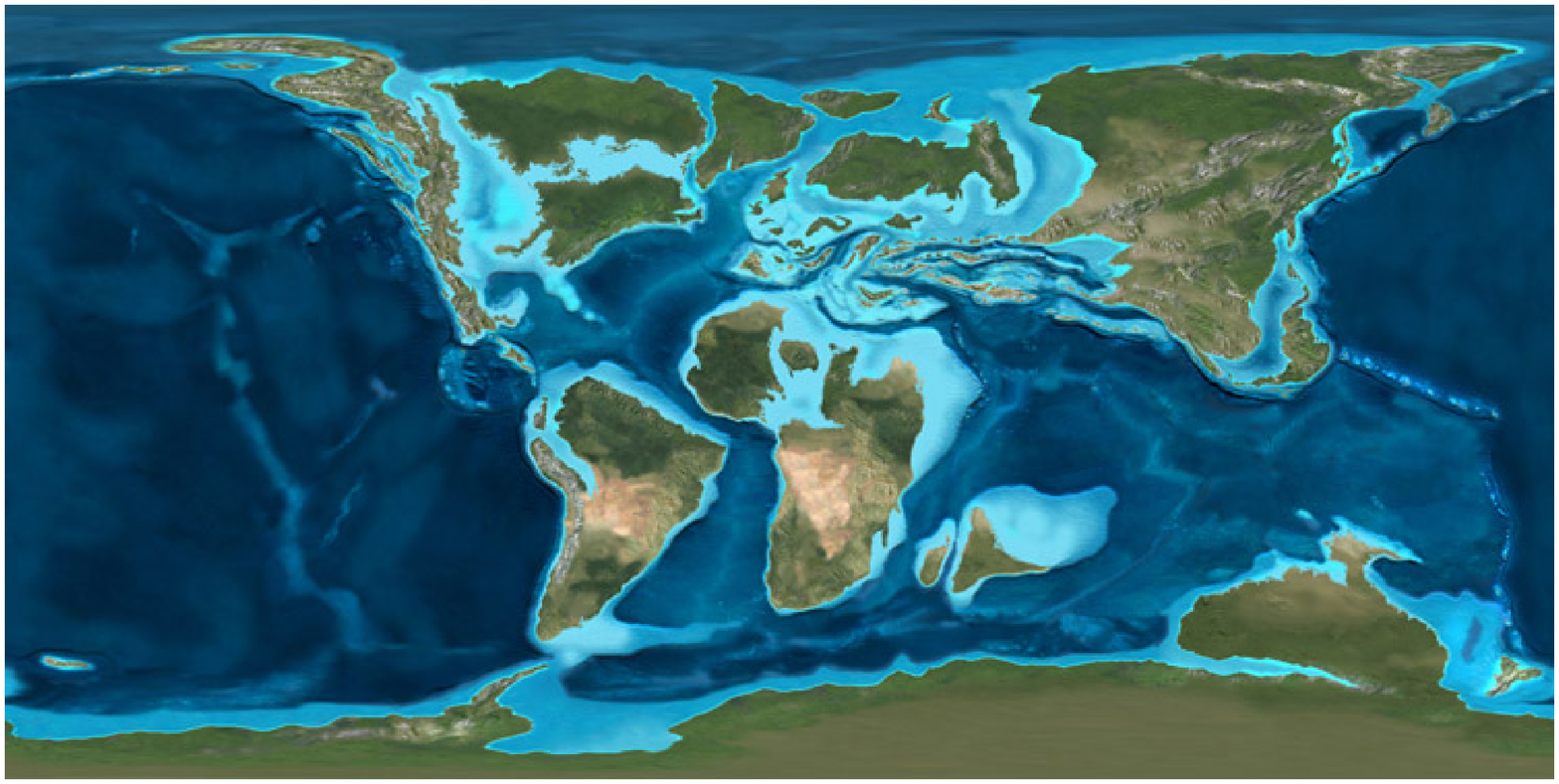}%
 \includegraphics[width=3.3in]{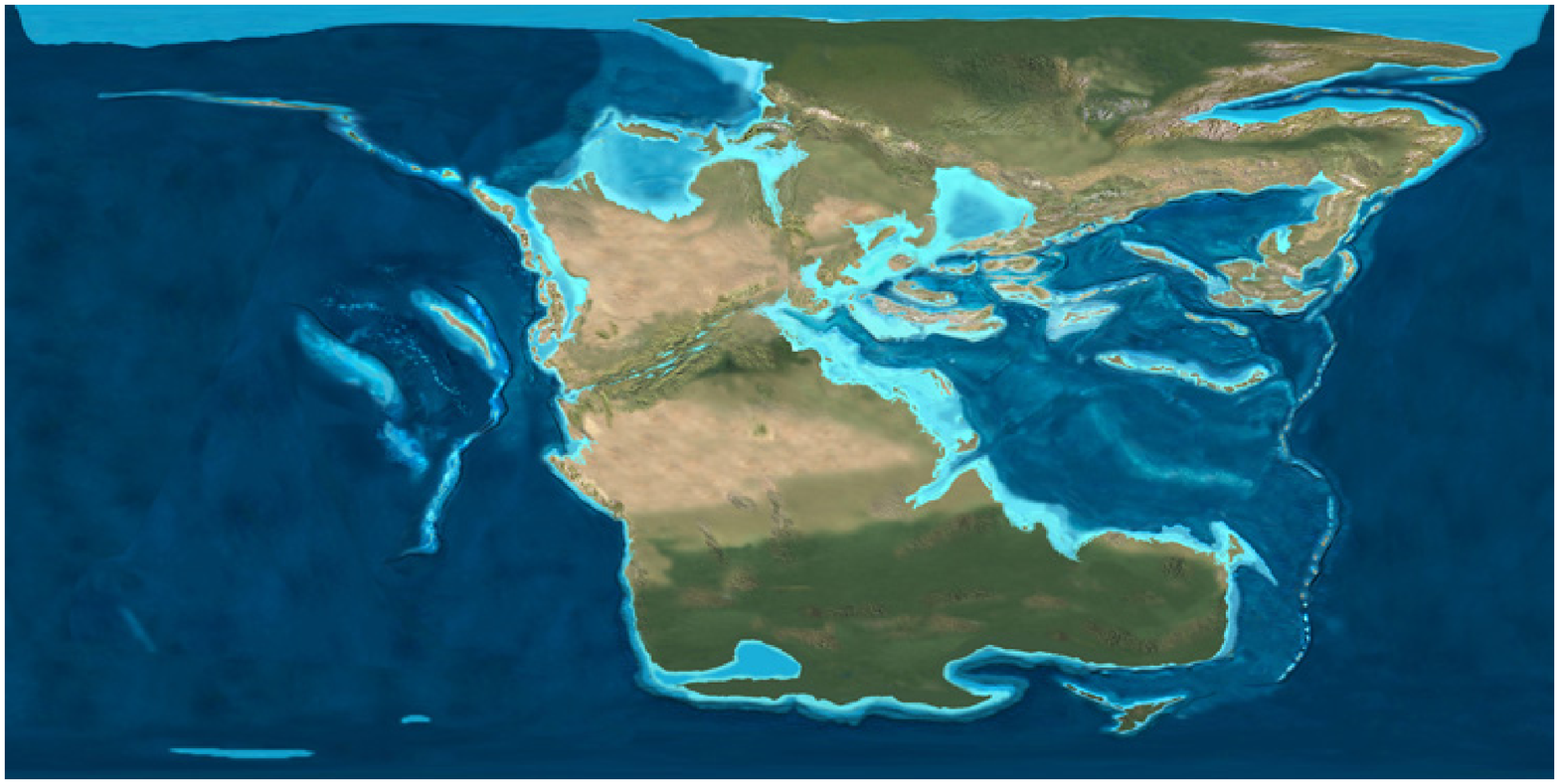}\\
 \includegraphics[width=3.3in]{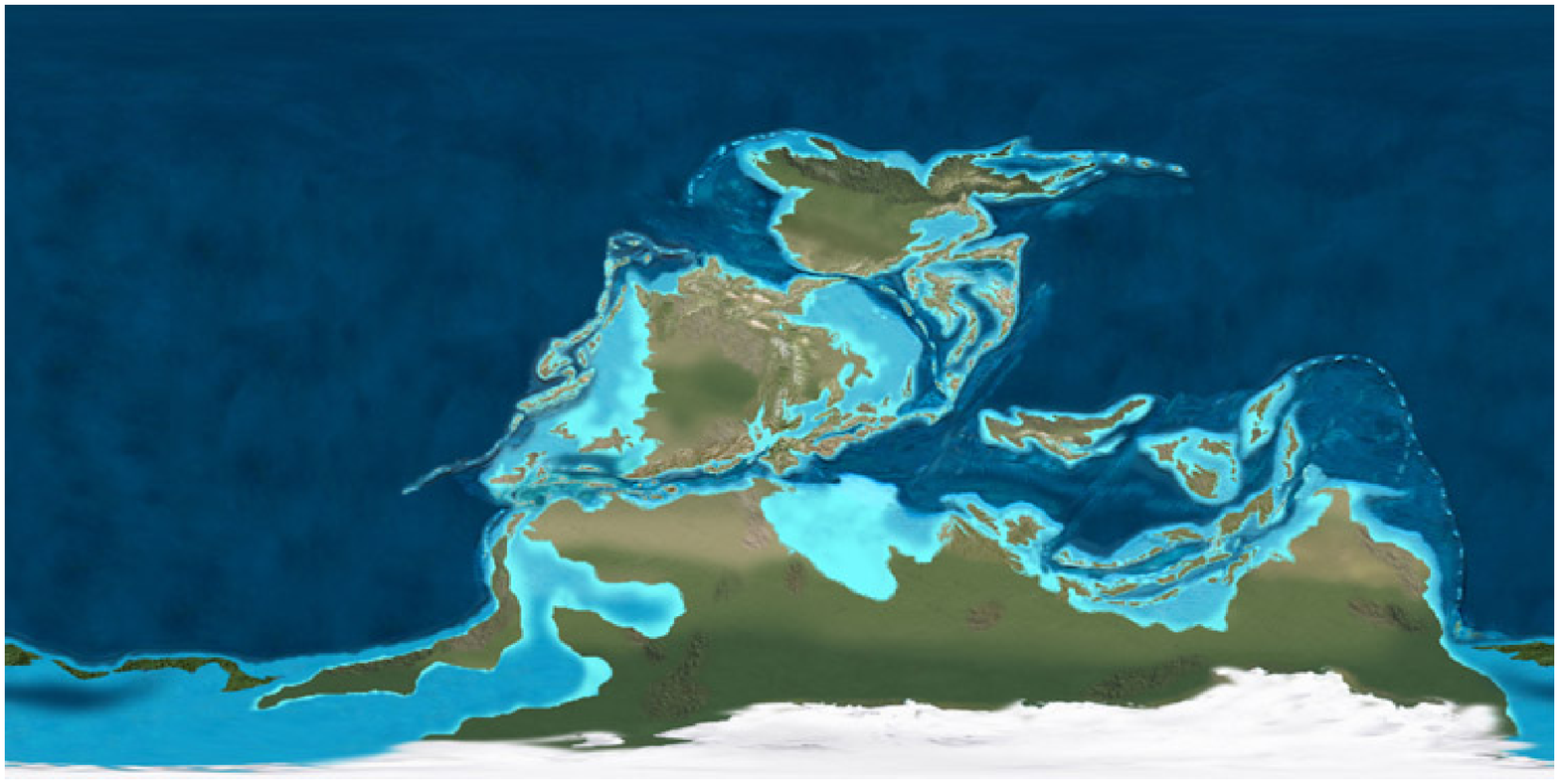}%
 \includegraphics[width=3.3in]{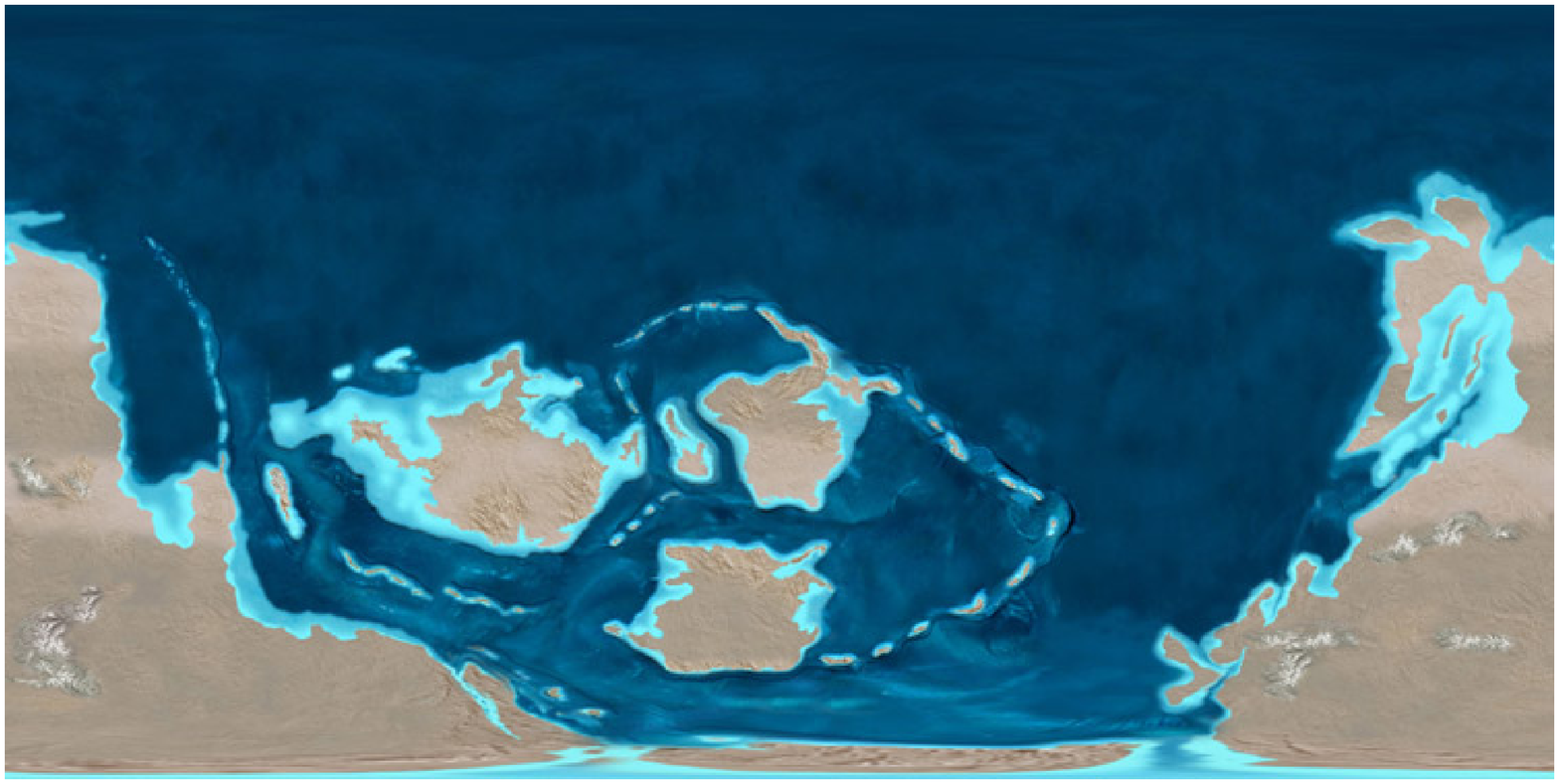}
\caption{Global views of the Earth's continental distribution during the Late Cretaceous (90 Ma ago; top left),
 the Late Triassic (230 Ma ago; top right), the Mississippian (340 Ma ago; bottom left), and the Late Cambrian (500 Ma ago; bottom right). Courtesy: Ron Blakey, Colorado Plateau Geosystems Inc.}
\label{past_epochs_maps}
\end{center}
\end{figure*}

In order to classify the correspondences between the land areas of these paleomaps and the ISCCP land types classification scheme,
we used an intensity thresholding method \citep{Son93}. Next, the information was regridded to a geographical resolution of 144 
(longitude) by 72 (latitude) cells, equal to the ISCCP cloud products, as input for our albedo models. Figure \ref{past_epochs_maps} 
shows the four historical periods that we have chosen to carry out our study. These epochs are 90, 230, 340 and 500 
Ma ago. Note that we have assumed
that the Earth 500 Ma ago was entirely desert as the development of advance plants is believed to have taken place in the 
Late Ordovician (450 Ma ago) \citep{Gra85}, although fungi, algae, and lichens might have greened many land areas 
some time before. Whether this is really the case for the Late Cambrian or not is not the real issue, as our 
supposition is meant to illustrate the albedo properties for a planet with a continental crust but not 
life in its surface, which was 
probably the case for Earth in many previous epochs, and a plausible scenario for Earth-like extrasolar planets.

The selection of these four geological epochs corresponds to a period of time during which plate tectonics 
has radically changed the face of the planet. However, albeit with large temporal excursions, Earth's atmosphere can be considered similar to the present \citep{Har78,Kas02}. On average the same atmospheric composition and mean averaged temperature have existed during this period. This is a necessary condition for our supposition of cloud distribution related to latitude and surface type to hold as a valid approximation.

\section{A SEMI EMPIRICAL MODEL FOR CLOUDS}



When attempting to study the reflectance properties of the Earth in the past, one of the major difficulties is the complete
lack of reliable information on cloudiness at these time scales. One of the possibilities would be to model the expected
cloud amount based on global climate models tuned to those past epochs. However, this is not easy to do with reliability,
as in fact cloud amount and variability poses one of the most complicated puzzles for climate change today.
For example, despite results of numerous general circulation models, it is unclear how increasing atmospheric temperatures resulting
from anthropogenic forcings may influence global cloud properties \citep{Ces96}, which is important because due to the large radiative influence exerted
by cloud cover, a small change in cloud amount or distribution may potentially provide a strong feedback effect,
significantly enhancing or mitigating the effects of global warming \citep{Des10,Spe10}.
Here, we use a different approach to tackle this problem, by using a semi empirical model to map how clouds behave over Earth's 
surface depending on latitude and surface types - such as ice, water, desert, and vegetation - which then we apply to past epochs.

To reach this objective the first step has been to calculate the 1984-2006 climatology of ISSCP cloudiness data. With this 
climatology, we have performed a classification of the amount of clouds depending on surface types. That is, cloudiness data have 
been split up into four groups according to the surface type over which the cloud is located. To carry out this classification we 
have made use of real geographical information about the different kinds of surfaces and vegetation in our planet. The ISCCP 
classifies the different surfaces as water, rain forest, deciduous forest, evergreen forest, grassland, tundra, shrub-land, 
desert, and ice, but for our study we have only distinguished between water, desert, ice, and vegetation, where the latest is 
defined as such zones where there were rain forest, deciduous forest, evergreen forest, grassland, tundra, or shrub-land (separate analysis for these subgroups gave very similar results).

\begin{figure*}[b]
\begin{center}
 \includegraphics[width=6.3in]{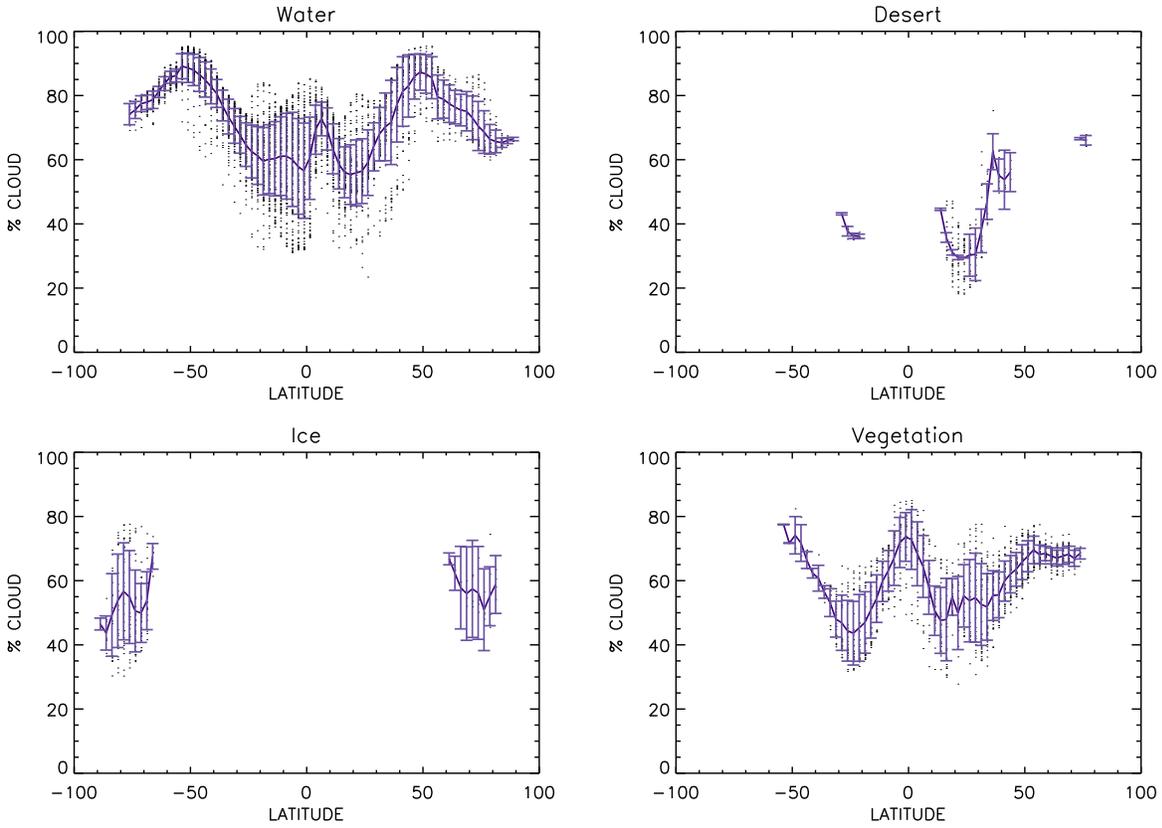}
\caption{Empirical relationships between the amount of clouds, surface type, and latitude. These figures show the ISCCP
1984-2006 average climatology of cloudiness for each grid cell on Earth as a function of latitude (dots), for each surface type 
separately: water (top left),
desert (top right), ice (bottom left), and vegetation (bottom right). Solid lines represent the mean cloudiness
at each latitude and the error bars represent the standard deviation of the mean.}
\label{empirical_relationships}
\end{center}
\end{figure*}

Once we got the classification of the cloudiness into these four groups, we represented the fraction of clouds
as a function of latitude for each surface type. We have used this scheme
to calculate the mean cloudiness at each latitude point and its corresponding standard deviation, thus obtaining empirical 
relationships between the amount of clouds, surface type and latitude. As can be seen in Figure \ref{empirical_relationships}, 
these derived cloudiness functions have a particular shape which is different for each surface type, suggesting that global 
cloudiness distribution can be empirically traced according to the latitude and the underlying surface type.
We have repeated this analysis for separate years and in seasonal and monthly climatologies with very similar results, most 
of them not shown in this paper for space reasons.

In the next step, these relationships are used in an attempt to reconstruct the global mean cloudiness 
of the present Earth and the possible cloud distribution of past epochs of Earth's history. In order to do that, 
we have taken each surface map, both for present and past epochs of the Earth (those calculated in Section 2), and 
we have assigned to each gridpoint a cloudiness fraction by using the information of the previously derived cloudiness 
relationships, taking into account both surface type and latitude. In this way we obtained global mean reconstruction 
of cloudiness for each surface map.

As can be seen in Figure~\ref{empirical_relationships}, because of the continental distribution of our own planet, we only have information about the fraction of clouds located over vegetation in the latitude range [-60\hbox{$^{\circ}$}, 80\hbox{$^{\circ}$}]. Thus in order to reconstruct the cloudiness of historical epochs with vegetated areas out of this latitude range, we would have to extrapolate our functions to get the corresponding cloudiness. Since data obtained from extrapolation are subject to greater uncertainty, we decided to take the cloudiness value corresponding to the higher available latitude of the Northern Hemisphere and assigned it as the fraction of clouds of the remaining Northern latitudes for which we had no information. For the Southern Hemisphere latitudes for which we had no information about cloudiness, we assigned to these the fraction of clouds corresponding to the same latitudes in the Northern Hemisphere. This filling process does not affect 
substantially our results because the polar regions have a low contribution to the total albedo.

A similar problem arises with clouds located over deserts (see Figure~\ref{empirical_relationships}).
As deserts on the present day Earth are mainly located at low latitudes, in order to avoid
extrapolations we have chosen to use only two cloudiness values. That is, we calculated the mean cloudiness over 
deserts in the latitude range [0\hbox{$^{\circ}$} - 30\hbox{$^{\circ}$}] and in the range [30\hbox{$^{\circ}$} - 90\hbox{$^{\circ}$}] and 
we assigned these mean values to those points which were located in the respective latitude range. 
These values are $33$\% for the former range and $55$\% for the last one.
Note that we have only used this modification in our empirical cloud reconstruction for the Cambrian (500 Ma ago), 
since for the other epochs desert areas are located mainly at low latitudes.

In the top left panel of Figure \ref{cloud_alb_actual}, the real Earth cloud distribution is shown. In the middle left panel it is 
shown the global cloud distribution as reconstructed from our simple model. As can be seen in the figure, our semi-empirical 
model for clouds reproduces well the general features of Earth's cloudiness distribution. The latitudinal range 
40\hbox{$^{\circ}$}-75\hbox{$^{\circ}$}, both north and south, is quite well reproduced, as well as the differences in cloud 
amount between land and water at the same latitude. There are, however, some small-scale structures, mostly over tropical oceans, 
that our model does not reproduce, 
being the major differences located in the Pacific and in the Indian oceans. These differences are due to oceanic effects, 
such as surface temperature, trade winds, ocean currents, and regional meteorological phenomena such 
as ``El ni\~no''/``La ni\~na'', which can greatly influence the cloud formation. This is difficult to parameterize in a 
general model valid for any continental distribution.

However, with the aim of improving our result over the oceanic areas, we have tried to capture the longitudinal variations in cloud 
amount by introducing an oceanic longitudinal variation in our model. We observe that in all large ocean basins, although the 
effect is larger for the Pacific, clouds tend to accumulate over the eastern margin (cold currents) and have a minimum over the 
western margins (hot currents). Taking the Pacific as the rule, we have selected the region where the major differences take 
place, i.e., the region between -20\hbox{$^{\circ}$} and 0\hbox{$^{\circ}$} latitude and between -70\hbox{$^{\circ}$} 
and -180\hbox{$^{\circ}$} longitude, approximately. We calculated the mean cloudiness over this box as a function of longitude and 
fit it with a two-degree polynomial curve (a parabola). We have applied this parabolic effect to all tropical ocean basins to make 
a second order correction to our modeled cloud distributions. 
To do that, the parabola was normalized to the mean cloud amount and then the cloud amount in each ocean basin was multiplied by 
this normalized parabola. As mentioned previously, the same correction is applied to all ocean areas within -20\hbox{$^{\circ}$} 
to 0\hbox{$^{\circ}$}, although the effect is almost unnoticeable for small oceans.

Figure \ref{cloud_alb_actual} bottom left shows the reconstructed cloud distribution of the Earth today
after applying the longitudinal correction. As can be seen cloudiness at the Pacific area has a similar
aspect to that shown in the same area in the real cloud map. In general, a differential map between the real 
cloud amount and our reconstructed cloudiness clearly indicates that this second order correction greatly improves the 
prediction errors.

\section{LIGHT CURVE RECONSTRUCTIONS}

Once we have been able to reconstruct the cloud distribution for present and past epochs of the Earth, we want to transform this 
information into Bond albedo values to test the photometric variability of the Earth as seen from a distant observer. 
To this end, we have used a simple Earth reflectance model, which calculates the visible reflected light curves of our planet for a 
given day, by using as inputs ground scene models from the Earth Radiation Budget Experiment \citep{Sut88}, 
cloud and snow/ice cover maps, and surface maps as inputs (Pallé et al. 2003). With this information the model is able to determine 
the Bond albedo at each time of the day and in any particular direction. Nevertheless, the model can only be considered as a first-order approximation since it does not take into account any other climate parameters beyond snow and ice that might contribute to changes in Bond albedo. 
Further, the model contemplates only $12$ different surface scenes and $4$ cloudiness levels ($0-5$\%, $5-50$\%, $50-95$\%, and $95-100$\%).

Our model calculates first the albedo of each element of area of the planet's surface taking into account surface type, 
cloud amount, snow/ice cover, and solar zenith angle. Then the model performs the calculation of the Bond albedo in terms of the albedo of each element of area by integrating over all portions of the globe illuminated by the Sun. For more details see \citep{Pal03}.

Obviously both cloudiness and surface type play a role in the albedo variability and deconvolving the effects of
both components can be an impossible task for extrasolar planets. However, \cite{Pal08} demonstrated that if 
sufficient signal to noise can be obtained, it is possible to detect both the presence of continents and the clouds. 
In any case, our aim here is to characterize the photometric variability of both components combined.

\subsection{Present Day Earth}

\begin{figure*}[b]
\begin{center}

 \includegraphics[width=2.5in]{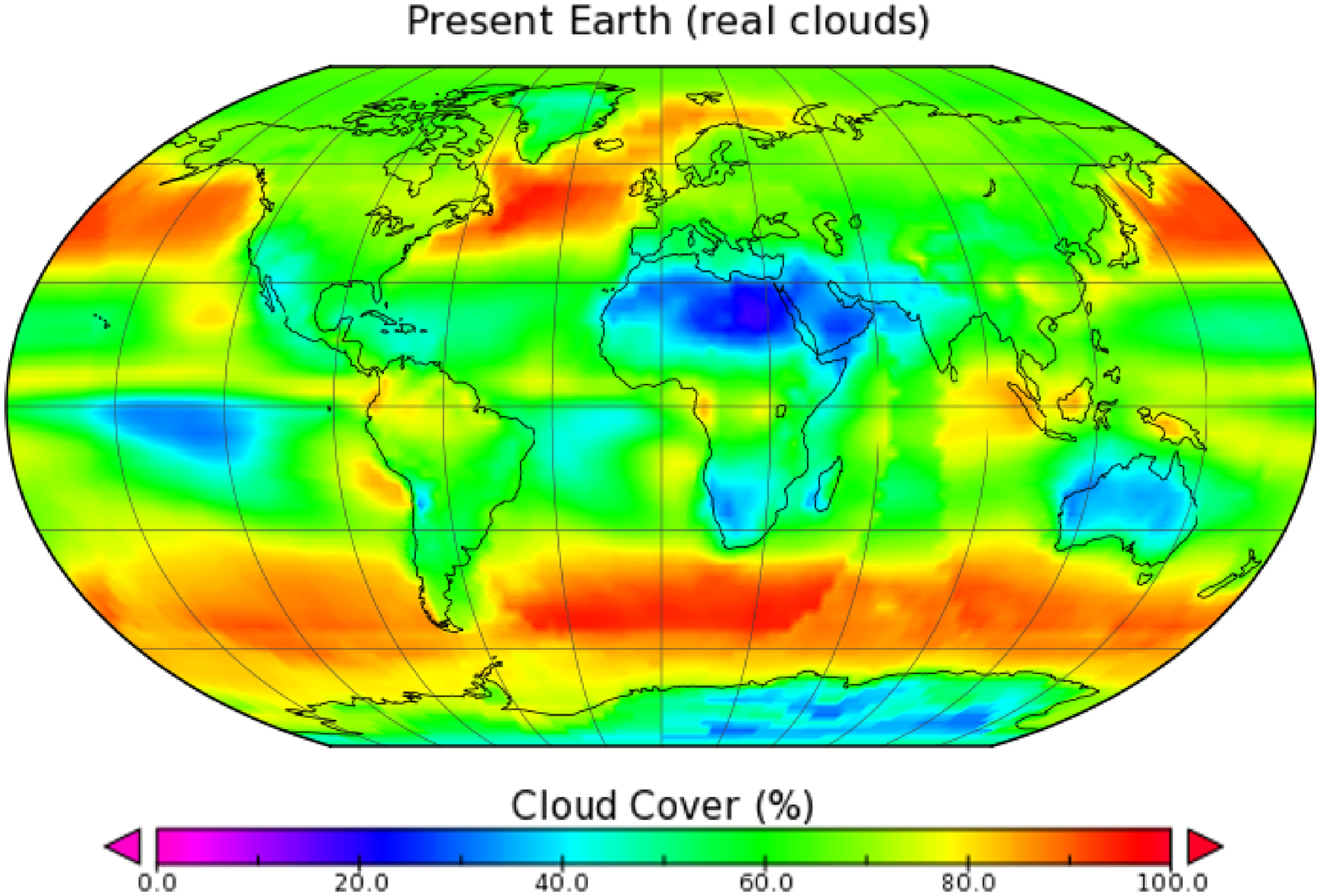}%
 \includegraphics[width=2.5in]{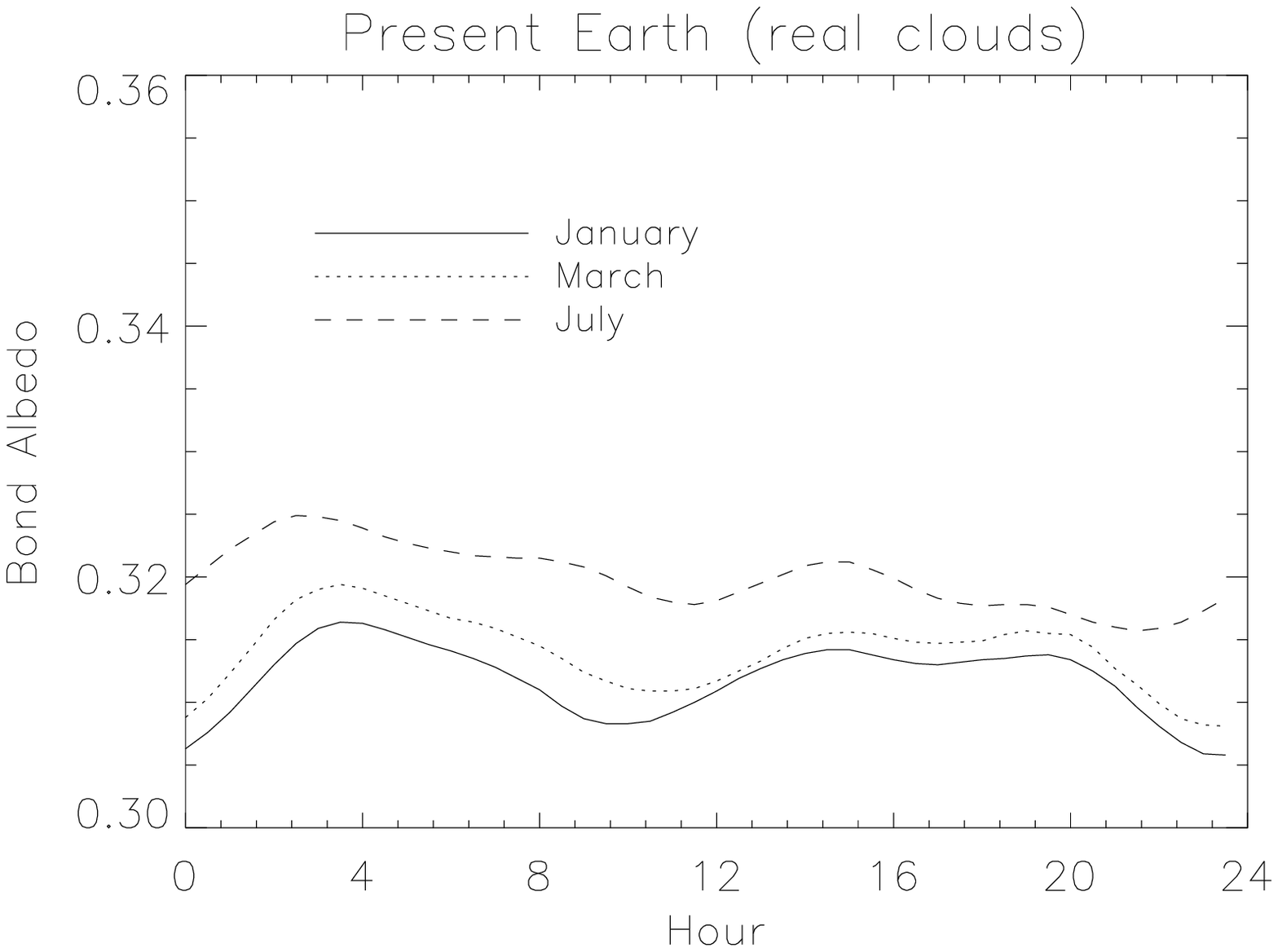}\\
 \includegraphics[width=2.5in]{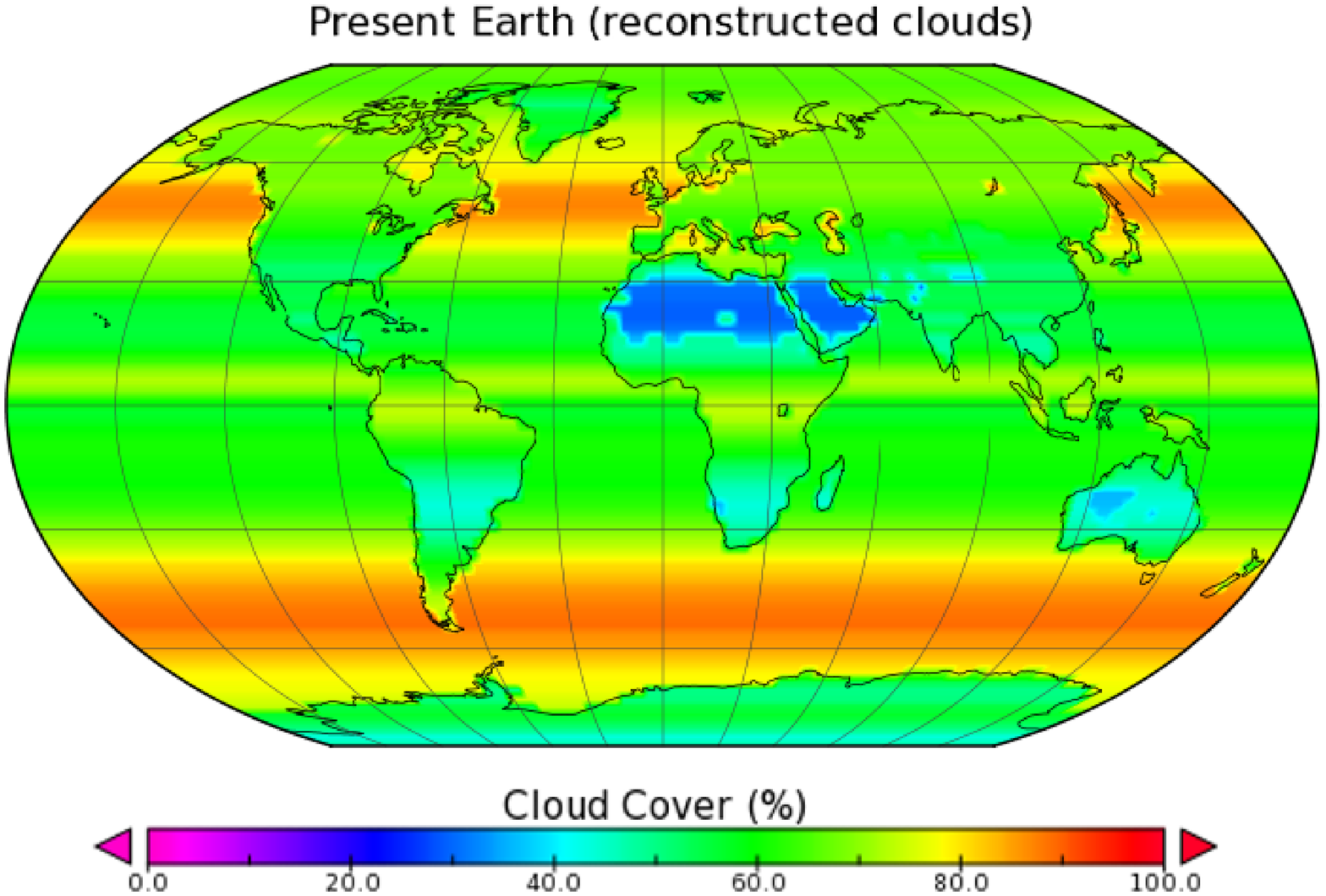}%
 \includegraphics[width=2.5in]{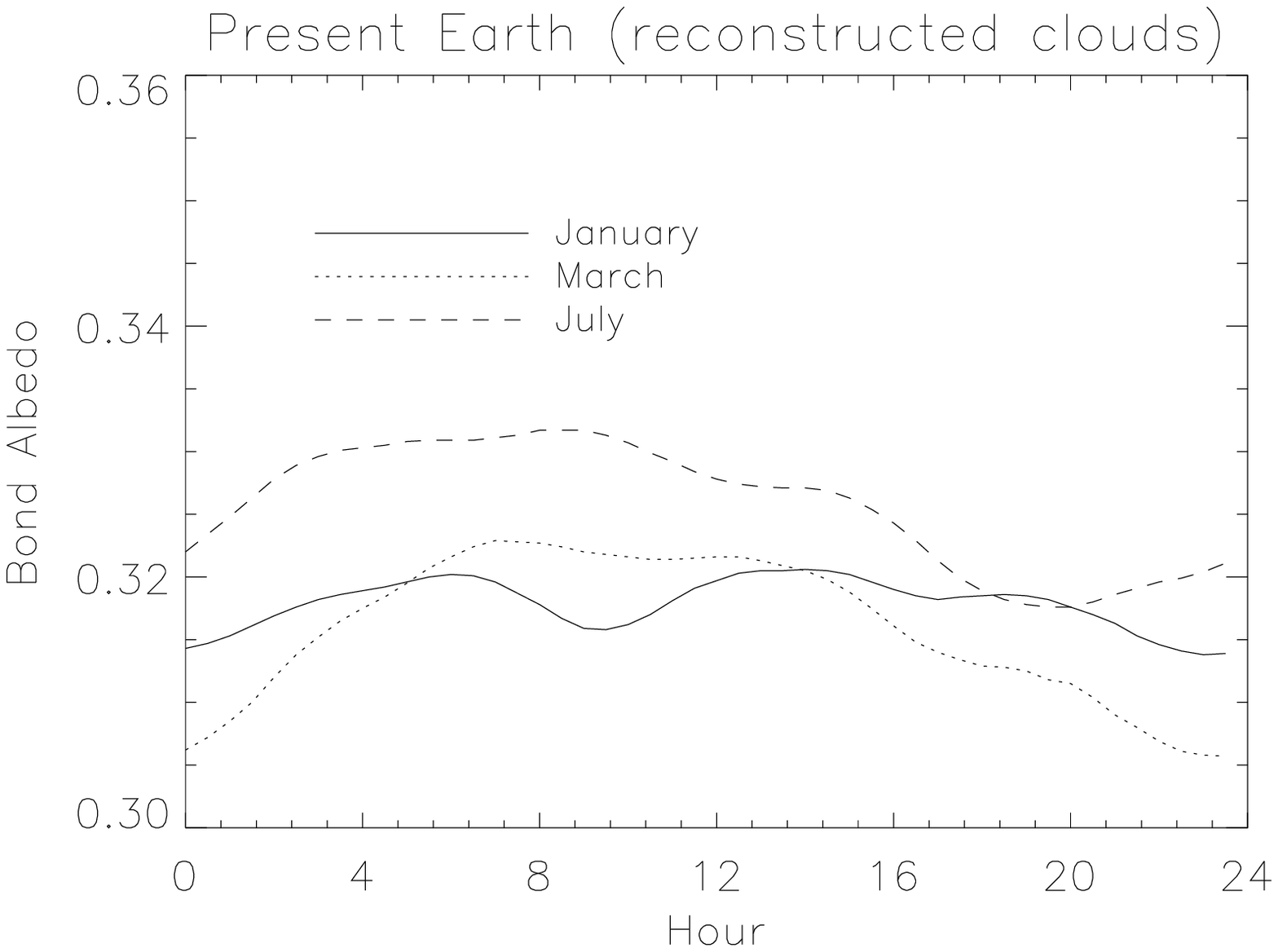}\\
 \includegraphics[width=2.5in]{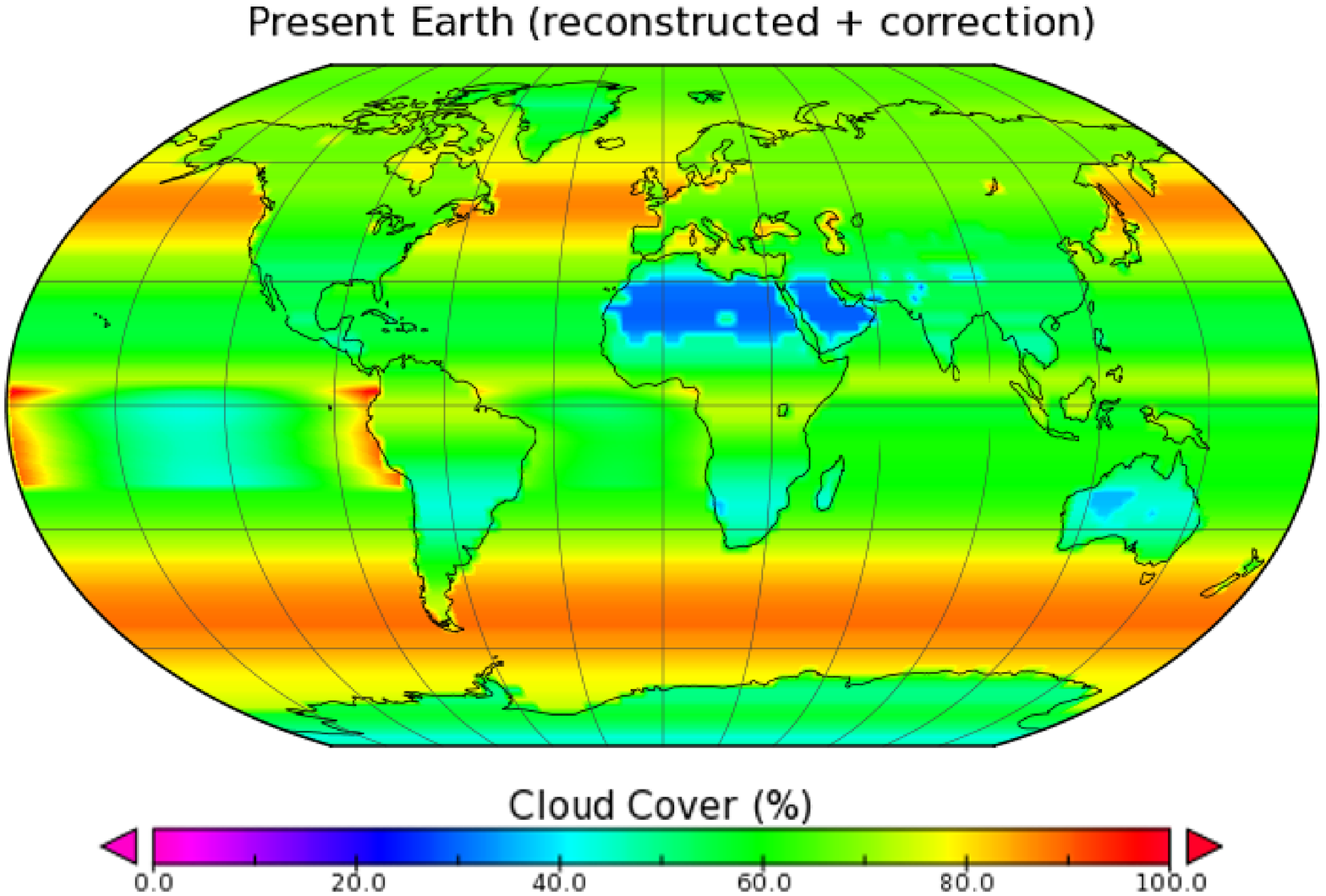}%
 \includegraphics[width=2.5in]{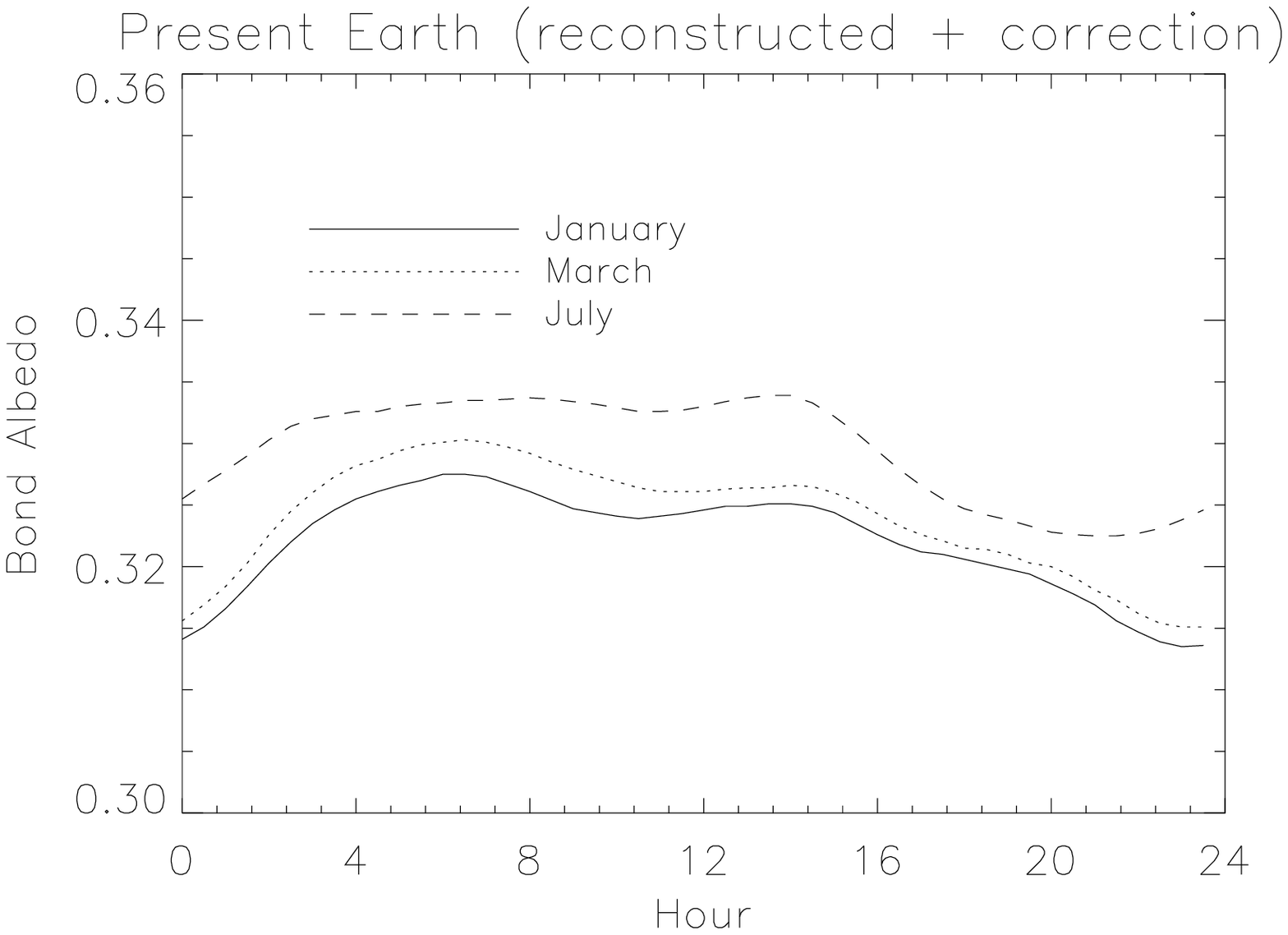}


\caption{Cloud distribution and light curves of the present Earth. Left panels illustrate the global distribution 
of real (top) and reconstructed (middle and bottom) cloud cover. Colors represent a different fraction of cloudiness as
the color code below each panel show. The difference between middle left and bottom left is that the last one
is the same as the former but after applying the parabolic correction over oceans (see the main text). 
Right panels show the corresponding simulations of Bond Albedo 24 hr variability. The different lines styles represent the 
results obtained for three different months as indicated in the legend.}

\label{cloud_alb_actual}
\end{center}
\end{figure*}

As representative light curves, we have arbitrarily chosen the months of January, March, and July $2000$ to carry out the Bond 
albedo simulations of the Earth at present day, and at 90, 230, 340, and 500 Ma ago. In the right panels of Figure 
\ref{cloud_alb_actual}, we plot the modeled 24 hr Bond albedo variation for the whole Earth over one day.

As can be seen in Figure \ref{cloud_alb_actual}, the agreement between the light curve obtained from reconstructed clouds 
(bottom right) and that calculated from real cloud cover (top right) is quite good.
The general shape of the light curves are quite similar, with the local maximums/minimums located at the same hour of the 
day in both cases. But the result obtained from the reconstructed cloudiness tend to be higher than that obtained from real 
cloud cover (0.325 as opposed to 0.315). Furthermore, the amplitude of the light curve is similar to that obtained from real 
cloud data (4.16\% and 3.29\% for reconstructed and real cloudiness, respectively). These differences in the variability of the 
Bond albedo are owing to the fact that real cloud cover presents larger changes in 
the longitudinal direction in contrast to the low variability in such direction of the 
reconstructed clouds, affecting Bond albedo values.

In Figure \ref{cloud_alb_actual}, it can also be seen that the resulting light curve after applying the longitudinal correction,
i.e., after applying the aforementioned parabolic smoothing technique applied over oceans, 
in the Pacific ocean (bottom-right) is more similar to that obtained with real cloud data, than
the light curve calculated before the correction (middle-right), confirming that the Earth cloudiness behavior is better 
reproduced by using this longitudinal correction.

\subsection{Historical Epochs}

\begin{figure*}[b]
\begin{center}

 \includegraphics[width=2.5in]{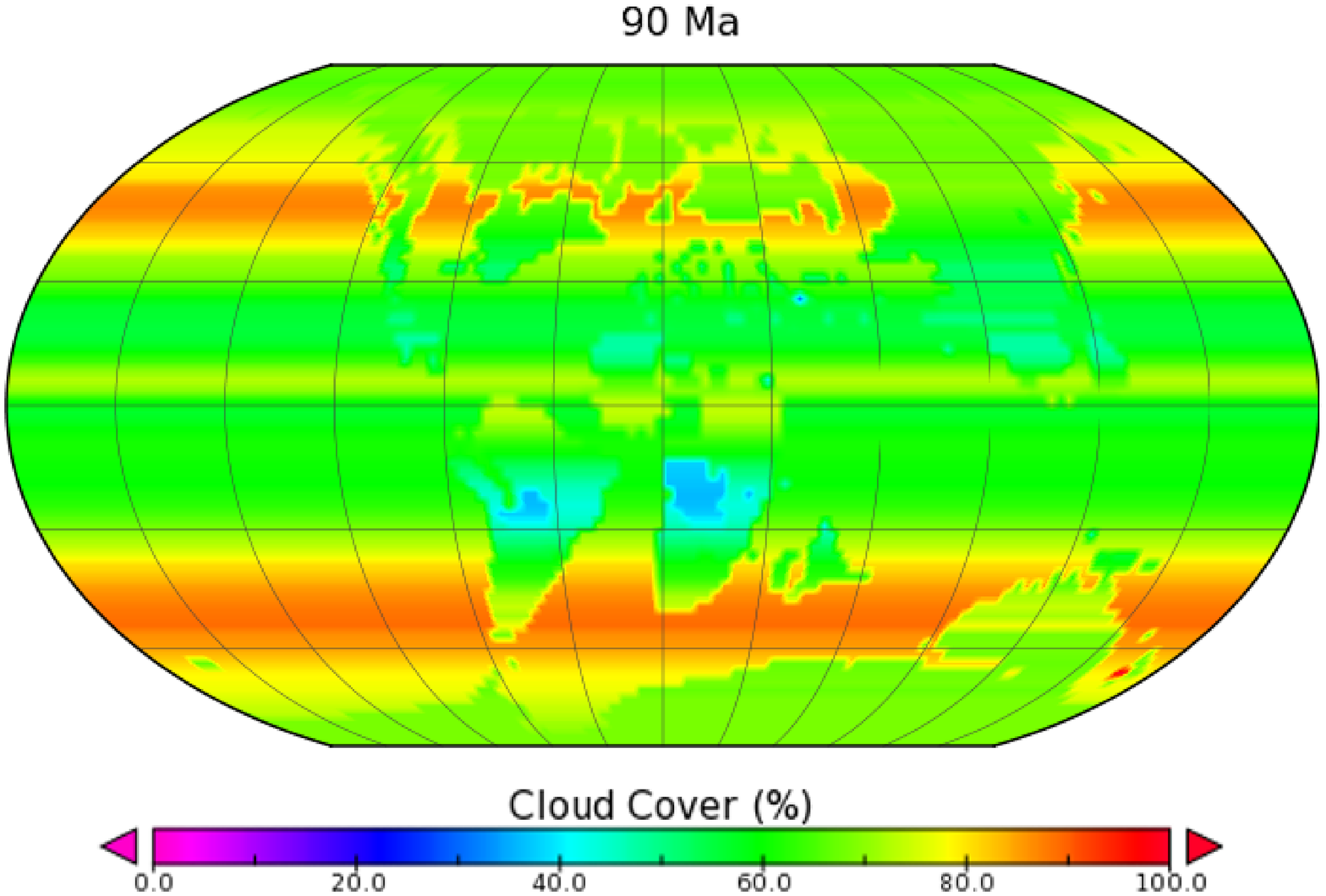}%
 \includegraphics[width=2.5in]{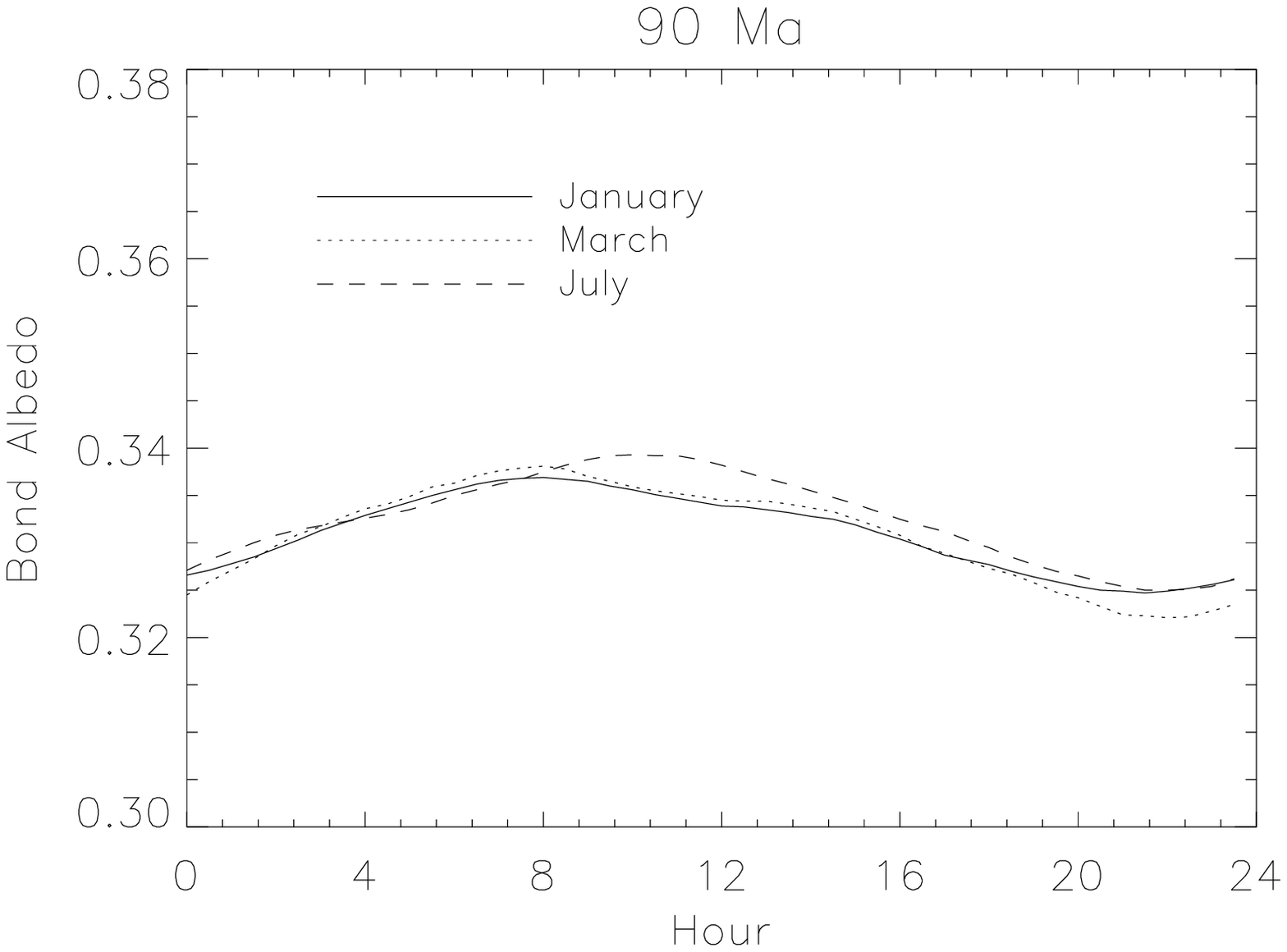}\\
 \includegraphics[width=2.5in]{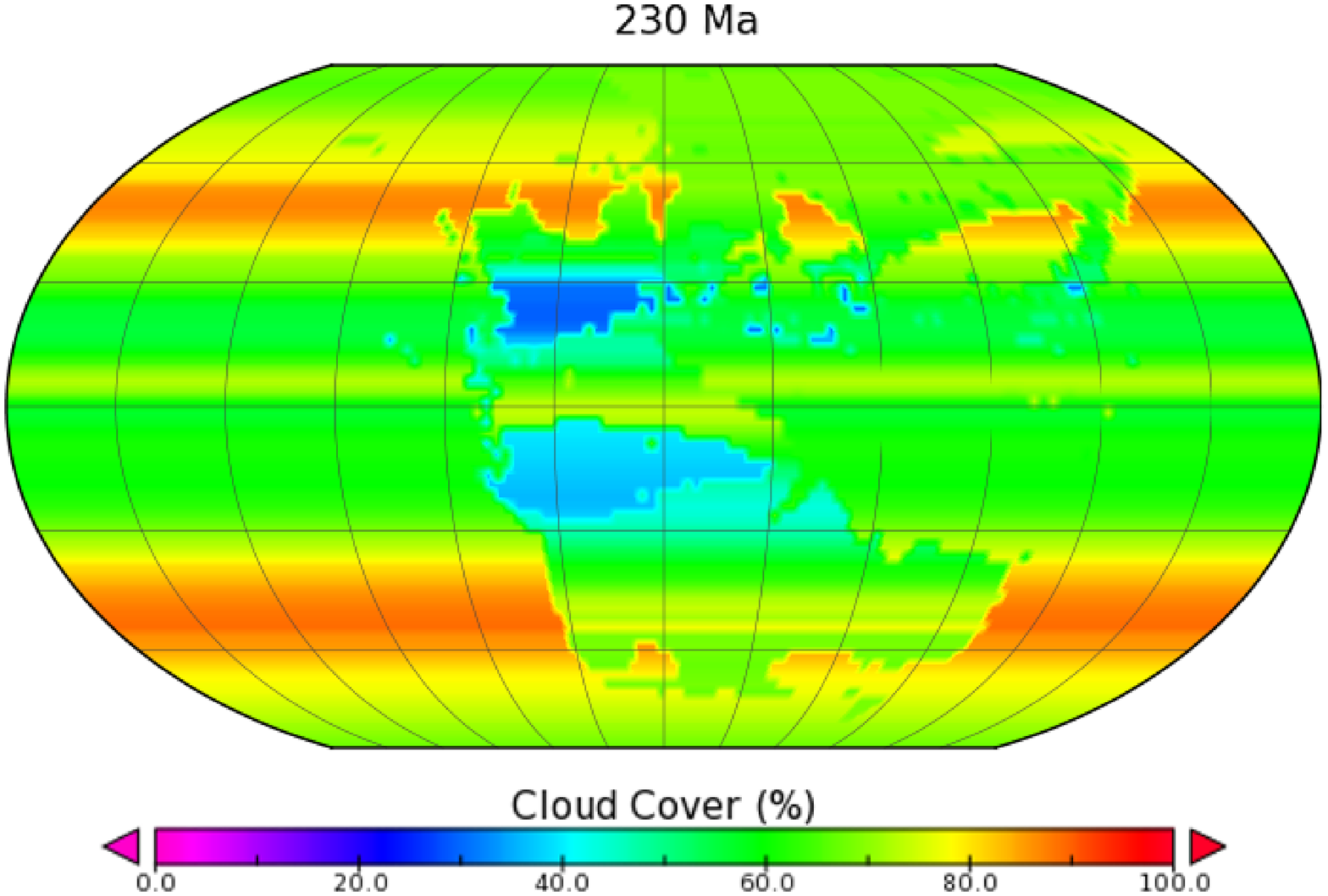}%
 \includegraphics[width=2.5in]{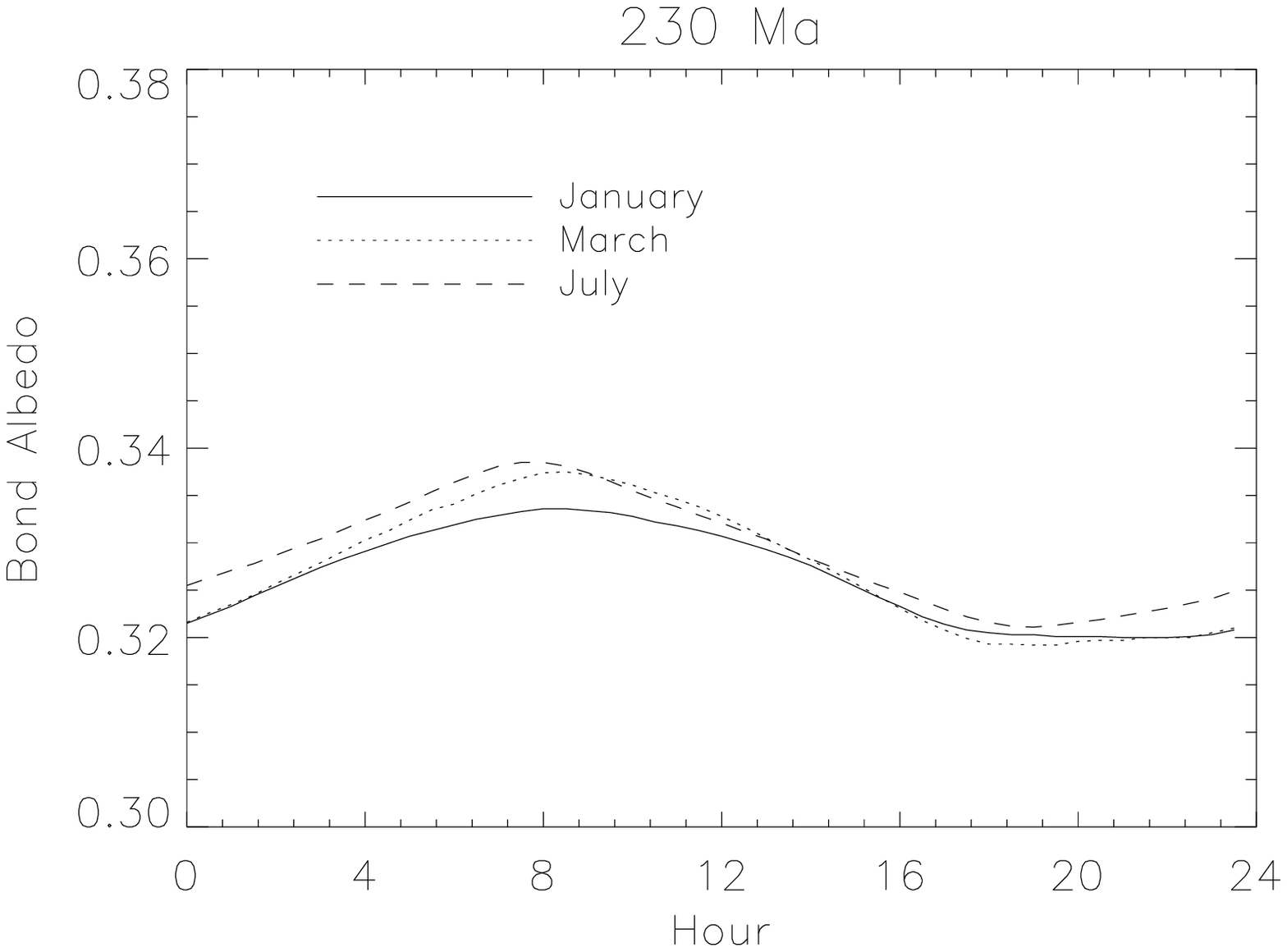}\\
 \includegraphics[width=2.5in]{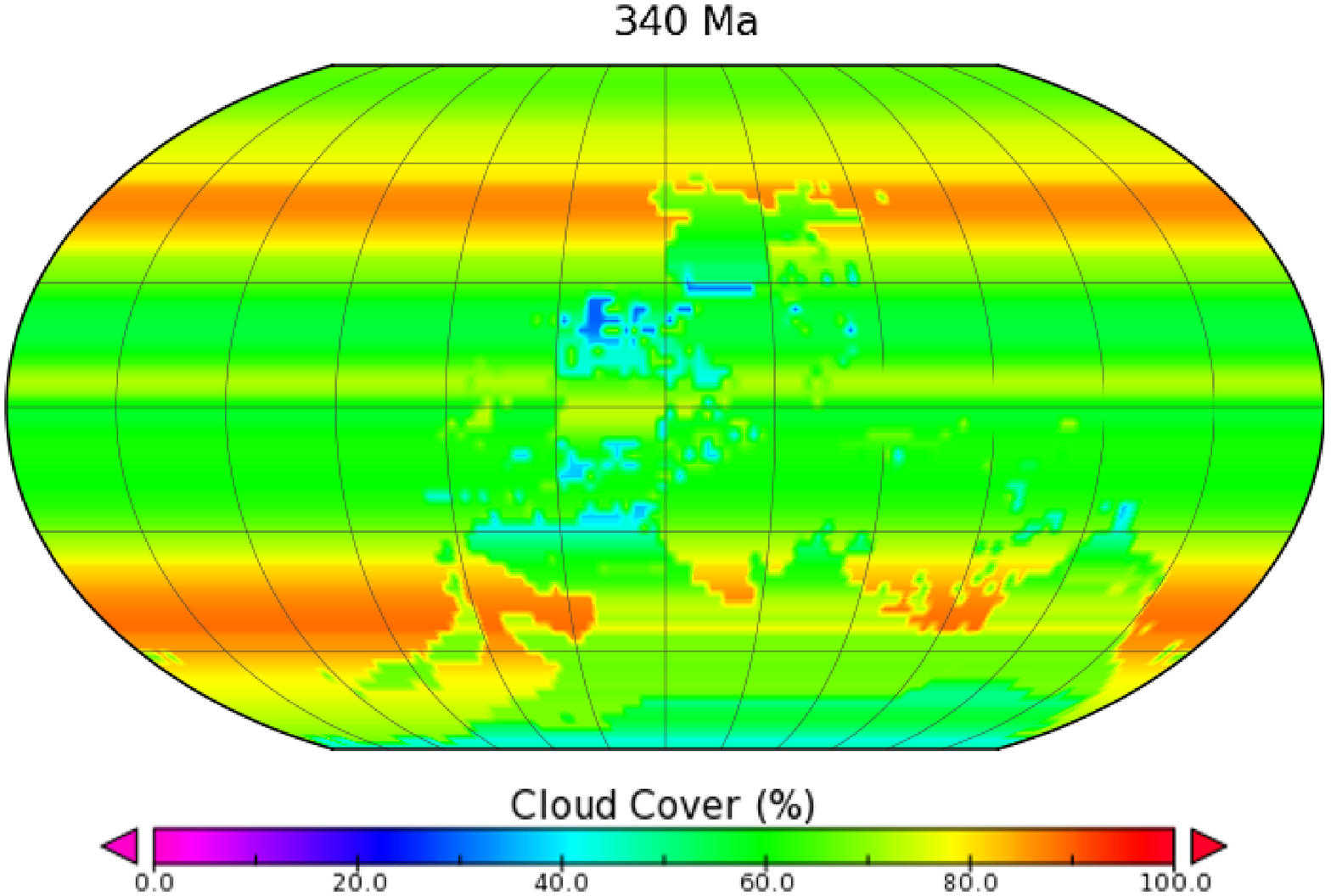}%
 \includegraphics[width=2.5in]{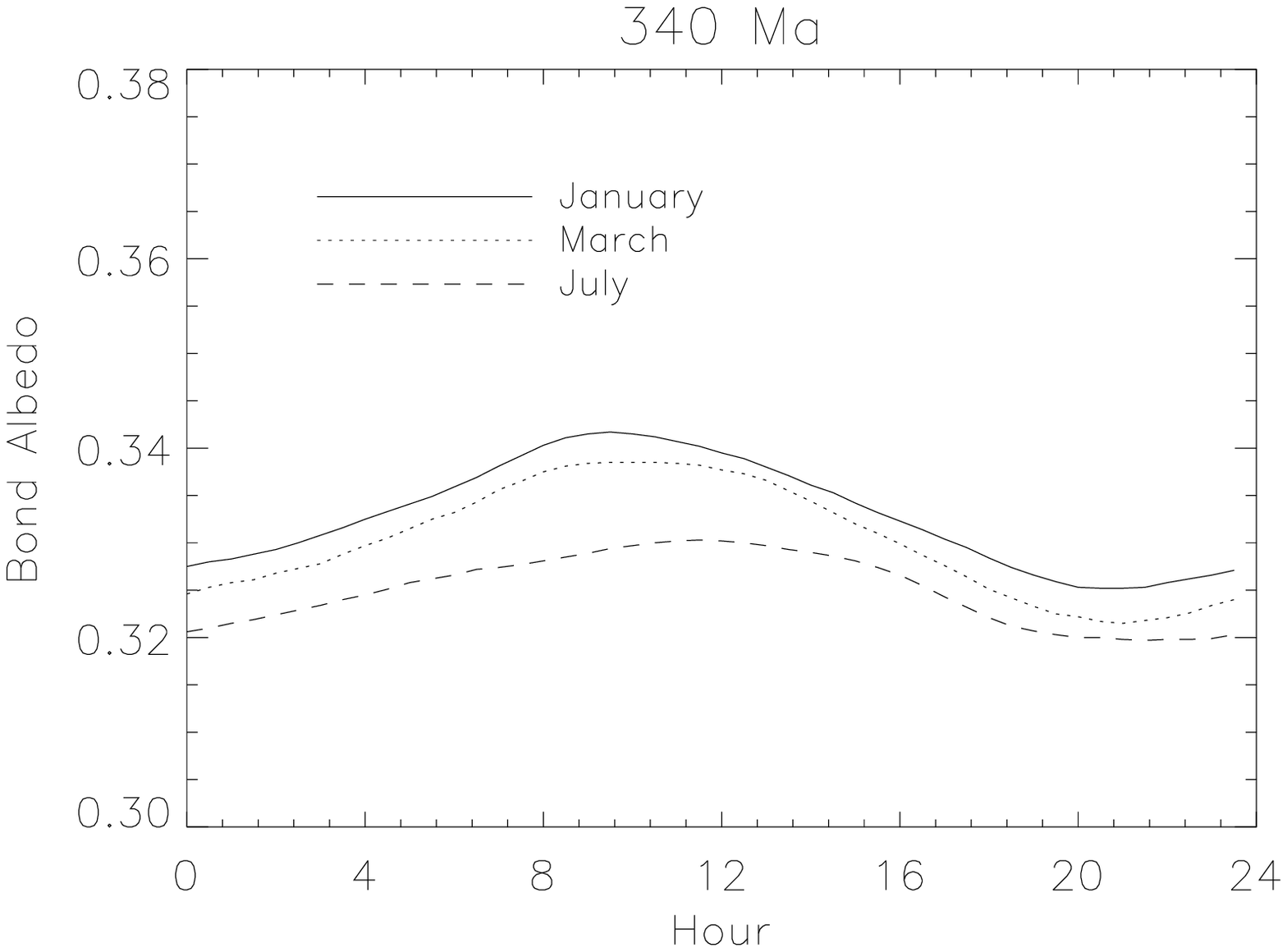}\\
 \includegraphics[width=2.5in]{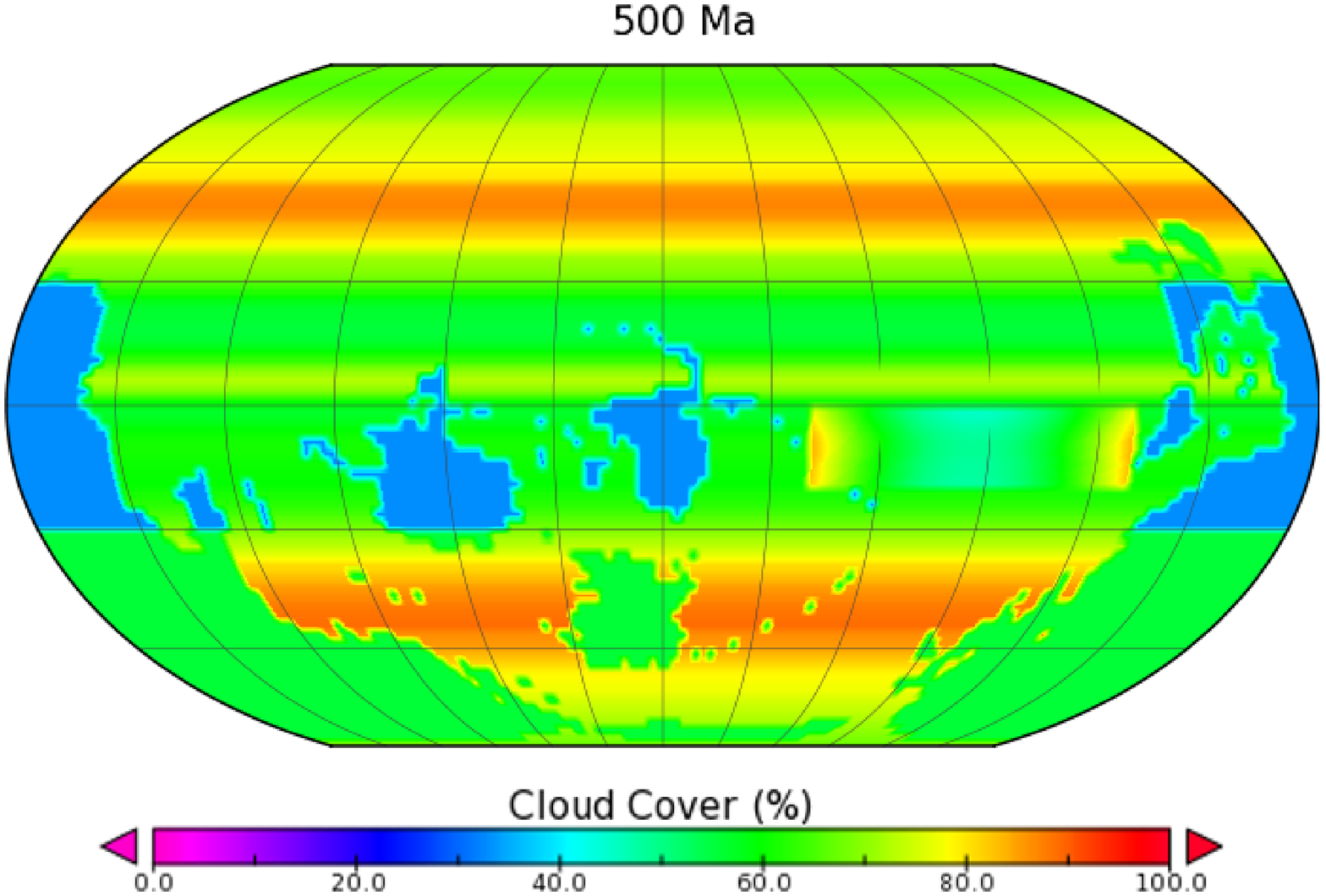}%
 \includegraphics[width=2.5in]{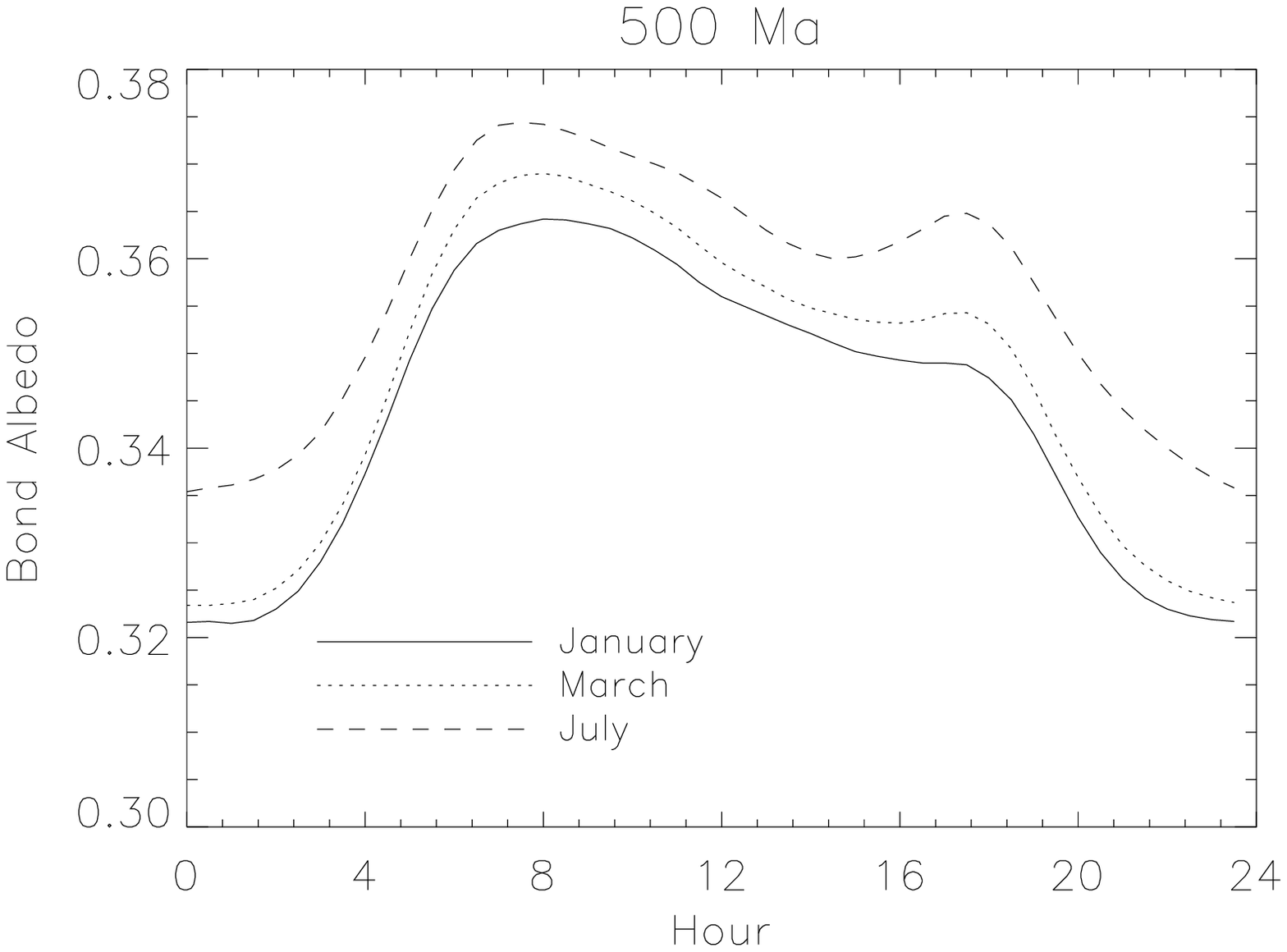}

\caption{Same as Figure\ref{cloud_alb_actual} but, from top to bottom, during the Late Cretaceous (90 Ma ago),
 the Late Triassic (230 Ma ago), the Mississippian (340 Ma ago), and the Late Cambrian (500 Ma ago).}
\label{cloud_albedo_past}
\end{center}
\end{figure*}

In the right panels of Figure \ref{cloud_albedo_past}, the light curves for the different epochs (90, 230, 340, and 500 Ma ago) are 
shown. As can be seen, the light curves of the first three epochs are quite similar. This is related to the fact that this epochs 
have a similar ocean-land-vegetation distribution, having a nearly totally clustered continents located principally in one hemisphere 
and deserts located at low latitudes. These light curves have a soft shape with an enhancement in brightness produced by continental 
masses. This smooth behavior is also due to the longitudinal relative homogeneity of the reconstructed cloud maps.

The reflected light of the Earth at 500 Ma ago, however, presents much more variability in contrast to the light curves obtained for 
the other epochs. That is again related to the continental distribution. In this epoch, most of the continents were clustered in one 
hemisphere, but it also has several big islands that cause strong variability in the reflected light. Moreover, this epoch presents a 
significantly higher mean albedo value. That can be related to the fact that in this epoch continents were covered by deserts, thus 
involving a higher reflectivity.

The mean Bond albedo values and their variability obtained for the Earth at different epochs are summarized in Table \ref{tbl-1}. By 
comparing the results shown in Table \ref{tbl-1} one can note that daily variations were larger than present day's, for instance 500 
Ma ago, when the variations were larger by a factor of three. Not only that, but also the mean albedo values were larger, which 
should have profound influences on the global climate by introducing a cooling effect, probably compensated by the increased 
greenhouse gas concentrations over that period, 16 times that of pre-industrial modern levels \citep{Aug04}.

\section{CONCLUSIONS}

In this paper, we have used information about the surface properties and continental distribution of the Earth with the aim of 
studying the behavior of the large-scale cloud patterns. We have obtained empirical relationships between the amount of cloudiness 
and their location on Earth's surface depending on latitude and underlying surface types. This relationships have been used to 
reconstruct the possible cloud distribution for different epochs of the Earth history such as 90, 230, 340, and 500 Ma ago. With 
this information, and with the help of an albedo model we have attempted, for the first time, to reconstruct and understand the 
photometric variability of these past epochs according to their different geographic and cloudiness distribution.

We find that our model reproduces well the major features of the cloud distribution and the photometric light curve of the Earth 
at present. When applied to past epochs of the Earth, we find that both the mean albedo value and the diurnal light curve 
variability remain stable as long as desert area are confined to the tropical regions. When this condition is not met, as 
during the Late Cambrian about 500 Ma ago, both the mean albedo value and the photometric variability are greatly increased. 
This increased variability could help in the determination of the rotational period of the planet from an astronomical distance. 
Due to the large compositional and chemical changes of Earth's atmosphere, we have not attempted to reconstruct cloud cover 
maps for epochs prior to the Late Cambrian. However, it is likely that the conditions for this period, i.e., higher albedo values 
and photometric variability, hold for much of the previous epochs of the Earth, not considering possible albedo variations due 
to atmospheric changes coming from clouds, aerosols, and hazes.

\acknowledgments

Research by E. Pall\'e is supported by a Ram\'on y Cajal fellowship from the MICIIN.




%


\begin{table}
\begin{center}
\caption{Percentage of the Daily Variability and Mean Albedo of the Light Curves for Each Epoch.
\label{tbl-1}}
\begin{tabular}{crrr}
\tableline\tableline
Epoch & $\%$ &  $\langle$Albedo$\rangle$ & \\
\tableline
Present Earth (real clouds) 		&3.29 &0.315\\
Present Earth (reconstructed clouds)    &4.16 &0.325\\
Late Cretaceous (90 Ma ago) 		&4.27 &0.331\\
Late Triassic (230 Ma ago) 		&5.02 &0.327\\
Mississippian (340 Ma ago) 		&4.46 &0.329\\
Late Cambrian (500 Ma ago) 		&12.2 &0.351\\
\tableline
\end{tabular}

\tablecomments{These quantities have been calculated by performing the mean of these results obtained
for 2000 January, March and July.}
\end{center}
\end{table}

\end{document}